\documentclass[
aps,
prd,
showpacs,
preprint,
tightenlines,
superscriptaddress,
nofootinbib,
floatfix]
{revtex4}
\usepackage{graphicx,
}
\begin{document}
%
\title{Three-flavor solar neutrino oscillations \\
with terrestrial neutrino constraints}
%
\author{        G.L.~Fogli}
\affiliation{   Dipartimento di Fisica
                and Sezione INFN di Bari\\
                Via Amendola 173, 70126 Bari, Italy\\}
\author{        G.~Lettera}
\affiliation{   Dipartimento di Fisica
                and Sezione INFN di Bari\\
                Via Amendola 173, 70126 Bari, Italy\\}
\author{        E.~Lisi}
\affiliation{   Dipartimento di Fisica
                and Sezione INFN di Bari\\
                Via Amendola 173, 70126 Bari, Italy\\}
\author{        A.~Marrone}
\affiliation{   Dipartimento di Fisica
                and Sezione INFN di Bari\\
                Via Amendola 173, 70126 Bari, Italy\\}
\author{        A.~Palazzo}
\affiliation{   Dipartimento di Fisica
                and Sezione INFN di Bari\\
                Via Amendola 173, 70126 Bari, Italy\\}
\author{        A.~Rotunno}
\affiliation{   Dipartimento di Fisica
                and Sezione INFN di Bari\\
                Via Amendola 173, 70126 Bari, Italy\\}

\begin{abstract}
We present an updated analysis of the current solar neutrino data
in terms of three-flavor oscillations,  including the additional
constraints coming from terrestrial neutrino oscillation searches
at the CHOOZ (reactor), Super-Kamiokande (atmospheric), and
KEK-to-Kamioka (accelerator) experiments. The best fit is reached
for the subcase of two-family mixing, and the additional admixture
with the third neutrino is severely limited. We discuss the
relevant features of the globally allowed regions in the
oscillation parameter space, as well as their impact on the
amplitude of possible CP-violation effects at future accelerator
experiments and on the reconstruction accuracy of the mass-mixing
oscillation parameters at the KamLAND reactor experiment.
\end{abstract}
\medskip
 \maketitle

\section{Introduction}

In this paper we present an updated three-flavor oscillation
analysis of the solar neutrino data coming from the Homestake
\cite{Cl98}, SAGE \cite{Ab02}, GALLEX/GNO \cite{Ha99,Ki02},
Super-Kamiokande (SK) \cite{Fu02}, and Sudbury Neutrino
Observatory (SNO) \cite{AhNC,AhDN} experiments. We include
additional constraints coming from terrestrial neutrino
oscillation searches performed at the CHOOZ reactors \cite{CHOO},
at the SK atmospheric $\nu$ experiment \cite{Shio}, and at the
KEK-to-Kamioka (K2K) long baseline accelerator experiment
\cite{K2Ke}. Implications for upcoming or future experiments are
also discussed. This work extends a previous solar $2\nu$ analysis
\cite{Getm}, to which we refer the reader for technical details.

The plan of our paper is as follows. In Sec.~II and III we briefly
review the theoretical and experimental input, respectively. In
Sec.~IV we present the results of our $3\nu$ oscillation analysis.
In Sec.~V we discuss some implications for CP violation searches
at future accelerator experiments. In Sec.~VI we analyze the
accuracy of parameter reconstruction in the  Kamioka Liquid
scintillator Anti Neutrino Detector (KamLAND) experiment at
reactors \cite{Shir}. We draw our conclusions in Sec.~VII.

\section{Theoretical $3\nu$ framework and approximations}

We consider standard $3\nu$ oscillations among flavor eigenstates
$\nu_\alpha=(\nu_e,\nu_\mu,\nu_\tau)$ and mass eigenstates
$\nu_i=(\nu_1,\nu_2,\nu_3)$ with a squared $\nu$ mass spectrum
defined as
\begin{equation}\label{masses}
(m^2_1,\,m^2_2,\,m^2_3) = \left(-\frac{\delta
m^2}{2},\,+\frac{\delta m^2}{2},\,\pm \Delta m^2\right)\ ,
\end{equation}
up to an irrelevant overall constant. The two squared mass gaps
$\delta m^2$ and $\Delta m^2$ (both $>0$) are usually referred to
as ``solar'' and ``atmospheric'' squared mass gaps, respectively.
The cases $+\Delta m^2$ and $-\Delta m^2$ identify the so-called
direct and inverted spectrum
hierarchies \cite{Qave}.%
\footnote{We assume direct hierarchy, unless otherwise stated. As
discussed later, the case of inverse hierarchy does not
appreciably change the results of our analysis.}

The mixing matrix $U_{\alpha i}$ is parametrized according to the
usual convention \cite{PDGR} as
\begin{equation}\label{mixings}
 U = U(\theta_{12},\theta_{13},\theta_{23})\ ,
\end{equation}
where we have dropped the (currently unobservable) CP violation
phase. The mixing angles span two octants, $\theta_{ij}\in
[0,\pi/2]$, and are often parametrized in terms of either
$\tan^2\theta_{ij}$ (in logarithmic scale) or $\sin^2\theta_{ij}$
(in linear scale) \cite{Scio,Matt}. We use the latter
representation in this work.%
\footnote{The alternative representation in terms of
$\log\tan^2\theta_{ij}$ is more useful when the allowed regions
span several decades in mixing angles---a situation no longer
realized in the current neutrino phenomenology.}

As it is well known, solar and terrestrial neutrino data
consistently favor the so-called hierarchical hypothesis
\begin{equation}\label{hier}
\delta m^2 \ll \Delta m^2\ ,
\end{equation}
which can be used to simplify the calculations \cite{Hier}. Even
if some regions of the explored parameter space do not fulfill the
above hypothesis, the {\em a posteriori\/} likelihood of combined
solutions in such regions appears to be so low that, at present,
the zeroth-order approximation (one-mass-scale dominance
\cite{Scio,Matt,Hier}) is practically justified in the current
analysis of both solar and atmospheric neutrino data.

In particular, solar neutrino oscillations are described with very
good accuracy in terms of the $3\nu$ parameter subset $(\delta
m^2,\theta_{12},\theta_{13})$ \cite{Matt,Pala,Cons}. Corrections
due to violations of Eq.~(\ref{hier}) have been shown to be
typically small (see \cite{Qave} and references therein), and can
be safely neglected at present. Therefore, in the statistical
analysis of the solar $\nu$ data, the $\chi^2$ function takes the
form:
\begin{equation}\label{chisolar}
\chi^2_{\mathrm{solar}} = \chi^2(\delta
m^2,\theta_{12},\theta_{13}) \ ,
\end{equation}
with no dependence on $\Delta m^2$ and thus on the mass spectrum
hierarchy (direct or inverse).

Concerning atmospheric neutrinos, it has been shown in several
analyses that the usual bounds on the dominant
$\nu_\mu\to\nu_\tau$ oscillation parameters $(\Delta
m^2,\theta_{23})$ are rather robust under perturbations beyond the
one-mass-scale dominance, induced by either $\theta_{13}> 0$
\cite{Subd,Shio,Vall,Yasu}
or by $\delta m^2> 0$ or both \cite{Cons,Stru,Tesh,Marr,Malt}
(within reactor bounds).%
\footnote{Moreover, the estimates of the leading parameters
$(\Delta m^2,\theta_{23})$ are robust under perturbations induced
by new neutrino states or interactions, see \cite{TAUP} and
references therein.}

Moreover, since the $\theta_{23}$ value is irrelevant for the
analysis of both solar and reactor neutrino data, we only need to
know the atmospheric  $\Delta m^2$ scale in our analysis, as
derived by a combination of the (preliminary) SK and K2K bounds
\cite{Shio,K2Ke} for unconstrained $\theta_{23}$ (see the next
section). Summarizing, we use the SK (atmospheric) and K2K bounds
to derive the marginal likelihood for $\Delta m^2$, namely,
\begin{equation}\label{chiSK+K2K}
\chi^2_{\mathrm{SK+K2K}} = \chi^2_{\mathrm{SK+K2K}}(\Delta m^2) \
.
\end{equation}
Such likelihood is practically independent of all the other $3\nu$
parameters (and on the $\pm \Delta m^2$ cases), as far as the
hierarchical hypothesis holds phenomenologically.

In the analysis of the CHOOZ reactor constraints, however, the
hierarchical approximation may not be accurate enough, and the
full $3\nu$ oscillation probability (as reported, e.g., in
\cite{Qave}) must be used in global analyses
\cite{Vall,Mont,Marr,Nico,Piai,Band}, providing a $\chi^2$
functional dependence of the kind
\begin{equation}\label{chiCHOOZ}
\chi^2_{\mathrm{CHOOZ}} = \chi^2_{\mathrm{CHOOZ}}(\delta
m^2,\pm\Delta m^2,\theta_{12},\theta_{13}) \ .
\end{equation}
The reason is that the CHOOZ bounds on $\overline\nu_e$
disappearance can be saturated by taking either $\theta_{13}>0$ or
$\delta m^2>0$  or both \cite{Marr,Nico,Moci,Malt}, thus providing
an anticorrelation between the CHOOZ upper limits on $\delta m^2$
and $\theta_{13}$  that must be properly taken into account. In
other words, the higher $\delta m^2$, the lower $\theta_{13}$, in
both cases of normal and inverted hierarchy \cite{Marr,Nico}. We
will discuss some consequences of such anticorrelation in Sec.~V.

Before performing the final combination of SK+K2K+CHOOZ with solar
neutrinos, it is useful to project away the functional dependence
of the likelihood on $\Delta m^2$ by defining a ``terrestrial''
$\chi^2$ as
\begin{equation}\label{chiterr}
\chi^2_\mathrm{terr}(\delta
m^2,\theta_{12},\theta_{13})=\min_{\Delta m^2} \left(
\chi^2_\mathrm{SK+K2K}+\chi^2_\mathrm{CHOOZ} \right)\ ,
\end{equation}
which contains only a residual dependence on the hierarchy
(direct or inverse) through $\chi^2_\mathrm{CHOOZ} $. Therefore,
in principle, the global combination
\begin{equation}\label{chiglobal}
\chi^2_\mathrm{global}(\delta m^2,\theta_{12},\theta_{13})=
\chi^2_\mathrm{solar}+\chi^2_\mathrm{terr}
\end{equation}
also depends on the hierarchy, although, unfortunately, in a very
weak way at present.

\section{Experimental input}

In this section we briefly review the experimental input coming
from solar and terrestrial (atmospheric, reactor, and accelerator)
experiments.

For the solar neutrino analysis we use the same standard solar
model input and neutrino data (81 observables) as in \cite{Getm},
but with an updated winter-summer rate difference from the
combination of GALLEX/GNO \cite{Ki02}%
\footnote{Corrected to remove eccentricity effects.}
and SAGE \cite{Ab02} measurements,
\begin{equation}\label{WS}
R_W-R_S \simeq  -5 \pm 7 \mathrm{\ SNU\ \ (GALLEX/GNO+SAGE)}\ .
\end{equation}

Concerning reactor data, we take the CHOOZ absolute spectrum from
\cite{CHOO}. The $3\nu$ analysis includes the 7+7 bin spectra as
in \cite{Qave}.

Concerning atmospheric neutrinos, the SK collaboration has
recently presented updated (92 kton yr) bounds on the leading
$\nu_\mu\to\nu_\tau$ atmospheric neutrino oscillation parameters
$(\Delta m^2,\sin^2 2\theta_{23})$ \cite{Shio}. Since the
correlation between the two parameters appears to be negligible,
the bounds on $\Delta m^2$ for unconstrained (i.e., projected)
values of $\sin^2 2\theta_{23}$ are basically obtained by fixing
$\sin^2 2\theta_{23}$ at its the best fit value (unity). From a
graphical reduction of the bounds shown in \cite{Shio}, we derive
then an approximate $\Delta\chi^2_\mathrm{SK}$ function in terms
of $\Delta m^2$.

The K2K Collaboration has also presented preliminary bounds on the
same mass-mixing parameters $(\Delta m^2,\sin^2 2\theta_{23})$
\cite{K2Ke}, again with apparently negligible correlation around
the best fit point (as far as $\sin^2 2\theta_{23}$ is nearly
maximal). The bounds on $\sin^2 2\theta_{23}$ are relatively weak,
but those on $\Delta m^2$ are already competitive with those
placed by SK and must be taken into account. We simply do so by a
graphical reduction of the $\Delta \chi^2_\mathrm{K2K}$ function
presented in \cite{K2Ke} for maximal mixing, to be combined
(summed) with the previous $\Delta \chi^2_\mathrm{SK}$.

It turns out that the function $\Delta \chi^2_\mathrm{SK+K2K}$
obtained in this way is nearly parabolic in the {\em linear\/}
variable $\Delta m^2$, although the separate
$\Delta\chi^2_\mathrm{SK}$ and $\Delta \chi^2_\mathrm{K2K}$
components are not parabolic. The SK+K2K $\chi^2$ minimum is at
$2.7\times 10^{-3}$ eV$^2$, with a $\pm 1\sigma$ error ($\Delta
\chi^2_\mathrm{SK+K2K}=1$ for $N_\mathrm{DF}=1$) equal to $\sim
0.4\times 10^{-3}$ eV$^2$. The estimated range for $\Delta m^2$,
as discussed in the previous section, is known to be very stable
under small perturbations induced by nonzero values of
$\theta_{13}$ or $\delta m^2$. Therefore, we use the following
SK+K2K combination,
\begin{equation}
\label{Dm} \Delta m^2 \simeq (2.7 \pm 0.4) \times 10^{-3}
\mathrm{\ eV}^2
\end{equation}
with approximately linear, symmetrical and gaussian errors, for
unconstrained values of the other $3\nu$ parameters. Notice that
values of $\Delta m^2$ below $1.5\times 10^{-3}$ eV$^2$ are
excluded at $3\sigma$, corroborating the trend of a previous
SK+K2K combination \cite{OurK}, and reinforcing the validity of
Eq.~(\ref{hier}). Of course, it will be desirable to confirm this
nice result by undertaking  a thorough and {\em ab initio\/}
analysis and combination of the K2K and SK data. We plan to do so
when such
preliminary data will be published and described in more detail.%
\footnote{In particular, the K2K {\em spectral shape\/} analysis
leading to the results in \cite{K2Ke} appears to involve much more
(and currently unpublished) information than it was required in
the simpler analysis of the {\em total rate\/} only \cite{OurK}.}

\section{Results}

In this section we describe the main results of our $3\nu$ solar
neutrino oscillation analysis with additional terrestrial
constraints. We start from the subcase of $2\nu$ oscillations.

\subsection{$2\nu$ oscillations}

In the subcase of pure $2\nu$ oscillations ($\theta_{13}=0$), the
information on $\Delta m^2$ coming from SK+K2K  is completely
decoupled within our approximations, and the solar+CHOOZ $\nu$
parameter space reduces to $(\delta m^2,\theta_{12})$.

Figure~1 shows the constraints coming from the global solar
neutrino analysis in the $(\delta m^2,\sin^2\theta_{12})$ plane,
with and without CHOOZ. In the case of solar $\nu$ data only, the
solutions are almost coincident with those presented in Fig.~1 of
\cite{Getm}, modulo the small changes induced by the updated input
from Eq.~(\ref{WS}) and by the different abscissa. For
completeness, Table~I gives the local $\chi^2$ minima for the
so-called large mixing angle (LMA), quasivacuum oscillation (QVO)
and low-$\delta m^2$ (LOW) solutions. These minima are not altered
by the inclusion of the CHOOZ data, whose only effect is to
slightly strengthen the upper bounds on $\delta m^2$, as shown in
Figure~1. The features of such $2\nu$ allowed regions have been
discussed in a number of papers
\cite{AhDN,Fu02,Ba02,Cr01,Ch02,Pe02,Ho02,St02} and are not
repeated here.

\begin{table}[t]
\caption{\label{tab2} Absolute and local best fits for the $2\nu$
analysis reported in Fig.~1 (without CHOOZ). The value
$\chi^2_\mathrm{min}=72.3$, reached in the LMA solution,
corresponds to a good overall fit to  81 solar neutrino data
(minus 2 free parameters).}
\begin{ruledtabular}
\begin{tabular}{lccc}
Solution & $\delta m^2$ (eV$^2$)& $\sin^2\theta_{12}$ & $\Delta\chi^2$  \\
\hline
LMA & $5.5\times 10^{-5}$ & $0.30$ & --  \\
QVO & $6.5\times 10^{-10}$ & $0.55$ & $7.5$   \\
LOW & $7.3\times 10^{-8}$ & $0.41$ & $10.5$
\end{tabular}
\end{ruledtabular}
\end{table}

\subsection{$3\nu$ oscillations}

Figure~2 shows sections of the volume allowed by solar neutrino
data only in the $3\nu$ parameter space $(\delta
m^2,\sin^2\theta_{12},\sin^2\theta_{13})$, for four representative
values of $\sin^2 \theta_{13}$ (equal to 0, 0.02, 0.04, and 0.06).
The best fit is reached in the same LMA point as for the $2\nu$
case $(\sin^2\theta_{13}=0)$. The constraints are derived through
Eq.~(\ref{chisolar})  and $N_\mathrm{DF}=3$. It can be seen that,
for increasing $\sin^2\theta_{13}$, the upper bounds on $\delta
m^2$ in the LMA become slightly weaker, and the LOW and QVO
solutions become less unlikely. Maximal ($\nu_1,\nu_2$) mixing
($\sin^2\theta_{12}=1/2$) is also less disfavored for increasing
$\sin^2\theta_{13}$. This trend is qualitatively consistent with
the recent results in \cite{Ho02} (obtained for an almost
equivalent solar $\nu$ data set), as well as with previous $3\nu$
analyses of solar neutrino data \cite{Pala,Vall,Band,YNir}.

Looser constraints on either $\delta m^2$ or maximal mixing can be
a desirable property from a theoretical or experimental point of
view. Relatively large values of $\delta m^2$ are, e.g., required
for the possible detection of leptonic CP violation in future
(very) long baseline experiments (see, e.g., \cite{CPre}). Nearly
maximal mixing can allow certain cancellations in $0\nu2\beta$
decay or in $\nu$ model building (see, e.g., \cite{Alta}).
Unfortunately, such ``desirable'' features in Fig.~2 are severely
limited by the inclusion of current terrestrial data through
Eqs.~(\ref{chiterr}) and (\ref{chiglobal}). This is shown in
Fig.~3, where the addition of the terrestrial $\nu$ constraints
actually strengthens the upper bounds on $\delta m^2$ for
increasing $\sin^2\theta_{13}$, since the CHOOZ bounds are more
quickly saturated. Notice also that, in comparison with earlier
analyses, the current K2K+SK bounds on $\Delta m^2$
[Eq.~(\ref{Dm})] are now sufficiently strong to prevent the
occurrence of relatively low values of $\Delta m^2$ ($\lesssim
1.5\times 10^{-3}$ eV$^2$), where the CHOOZ bounds would become
significantly weaker. Therefore, the current SK+K2K data enhance
the impact of terrestrial constraints on the solar $3\nu$
parameters $(\delta m^2,\sin^2\theta_{12},\sin^2\theta_{13})$.

Figure~4 provides another way of looking at the results of the
solar+terrestrial analysis. In this figure, the $\Delta\chi^2$
function is projected onto one parameter at a time, so that the
$n$-sigma bounds on each of them (the others being unconstrained)
are simply given by a $\Delta\chi^2=n^2$ cut. In the left panel it
can be seen that the $\chi^2$ difference between the LMA and the
LOW or QVO solutions is slightly lowered  in the $3\nu$ fit, as
compared with the $2\nu$ case reported in Table~I. However, such
improvement for the LOW and QVO solutions for nonzero
$\theta_{13}$  is currently marginal (less than one unit in
$\Delta\chi^2$). Basically all data converge  towards zero or low
values of $\sin^2\theta_{13}$, whose global upper bounds are
reported in the right panel of Fig.~4. In particular, one gets
\begin{equation}\label{phi}
\sin^2\theta_{13}<0.05\ (3\sigma,\;N_\mathrm{DF}=1)\ ,
\end{equation}
for unconstrained values of the other $3\nu$ parameters.

In conclusion, genuine $3\nu$ mixing in solar neutrinos is
severely constrained by terrestrial data. For the most favored
solution, increasing $\theta_{13}$ corresponds to lowering the
upper bound on $\delta m^2$. For the less favored LOW and QVO
solutions, the improvement of the fit for slightly nonzero values
of $\theta_{13}$ (see also \cite{Ho02}) is not statistically
significant. Figures~3 and 4 quantify such effects.

A final remark is in order. Figures~3 and 4 refer to the case of
direct neutrino spectrum hierarchy. We have verified that the
changes for inverse hierarchy (not shown) are negligible. Indeed,
the phenomenological difference between the two hierarchies, which
vanishes for $\theta_{13}\to 0$, is still too small to emerge for
the nonzero values of $\theta_{13}$ allowed by Eq.~(\ref{phi}).

\section{Implications for future CP violation searches}

The anticorrelation between the upper bounds on $\delta m^2$ and
$\sin^2\theta_{13}$ shown in Fig.~3 is not a particularly
desirable feature for future CP violation searches at
accelerators, where the difference between the $\nu$ and
$\overline \nu$ oscillation probabilities would be enhanced for
{\em both} $\delta m^2$ and $\theta_{13}$ close to their upper
limits (see, e.g., the reviews in \cite{CPre}).

In order to study in more detail the maximum CP violation effects
compatible with current bounds, we think it useful to focus on the
following benchmark quantity, which typically appears as an
overall prefactor modulating CP-violating amplitudes, and which
has the useful property to be a vacuum-matter (VM) invariant
\cite{Kimu}:
\begin{eqnarray}
I_\mathrm{VM}&=& (m^2_1-m^2_2)(m^2_2-m^2_3)(m^2_3-m^2_1)\sin
2\theta_{12} \sin 2\theta_{13} \cos \theta_{13} \\
&=& \delta m^2 \left[4(\Delta m^2)^2-(\delta m^2)^2\right]
\sin\theta_{12}\cos\theta_{12}\sin\theta_{13}\cos^2\theta_{13}\ .
\end{eqnarray}
More precisely, in order to eliminate the dependence on $\Delta
m^2$ and to simplify the discussion, we integrate $I_\mathrm{VM}$
over the gaussian distribution in $\Delta m^2$ defined in
Eq.~(\ref{Dm}), obtaining the expectation value
\begin{equation}
{\overline I}_\mathrm{VM} = \delta m^2 \left[4(\Delta
m^2_0)^2+4(\sigma_\Delta)^2-(\delta m^2)^2\right]
\sin\theta_{12}\cos\theta_{12}\sin\theta_{13}\cos^2\theta_{13}\ ,
\end{equation}
where $\Delta m^2_0=2.7\times 10^{-3}$ eV$^2$ and
$\sigma_\Delta=0.4 \times 10^{-3}$ eV$^2$ [Eq.~(\ref{Dm})]. The
quantity ${\overline I}_\mathrm{VM}$ is then a function of the
same $3\nu$ parameters $(\delta m^2,\theta_{12},\theta_{13})$ used
for the previous solar+terrestrial  analysis. Notice that
${\overline I}_\mathrm{VM}\to 0$ when any of these $3\nu$
parameters tends to zero, as intuitively expected for a quantity
associated to CP violation \cite{Kimu}.

Figure~5 shows isolines of ${\overline I}_\mathrm{VM}$ in the
plane $(\delta m^2,\sin^2\theta_{12})$, for the same
representative values of $\sin^2\theta_{13}$ as in Figs.~2 and 3.
The $\delta m^2$ scale is restricted to the range relevant for the
LMA solution, whose C.L.\ contours are superposed in each panel
(as taken from Fig.~3). In the upper left panel
($\sin^2\theta_{13}=0$), the vacuum-matter invariant ${\overline
I}_\mathrm{VM}$ is identically zero. In other panels, it can be
seen that ${\overline I}_\mathrm{VM}$ achieves its maximum values,
within the LMA solution, for ``intermediate'' values of
$\sin^2\theta_{13}$ ($\simeq 0.02$--0.04). Indeed, for increasing
values of $\sin^2\theta_{13}$ the LMA upper bound on $\delta m^2$
decreases, and ${\overline I}_\mathrm{VM}$ is maximized when a
compromise is reached. We conclude that the maximum allowed
amplitude of possible CP violation effects (in future accelerator
experiments) occurs somewhat below the upper bound placed by
terrestrial experiments on $\sin^2\theta_{13}$, due to the
phenomenological anticorrelation  with the $\delta m^2$ upper
bound.

\section{Implications for KamLAND}

The KamLAND experiment is currently detecting reactor
$\overline\nu_e$ events, whose energy spectrum  will provide soon
a clear terrestrial test of the solar $\nu$ LMA solution
\cite{Shir}. If the LMA solution is disconfirmed, the BOREXINO
experiment will be able to test the remaining LOW and QVO
solutions by detecting time variations of the event rate
\cite{BORE}. Assuming that the LMA solution is confirmed, various
studies have shown how well the KamLAND experiment can reconstruct
the oscillation parameters for  a few selected points in the
$(\delta m^2,\theta_{12})$ parameter space
\cite{Shir,Mura,Barg,BaSt,Gouv,GoPe,StVi,Alia}. In this section we
extend these studies by performing a continuous scan in the
$\delta m^2$ parameter, in the region of interest for the LMA
solution.

In the analysis, we use standard inputs for the reactor average
fuel composition \cite{Decl}, $\overline\nu_e$ energy spectra
\cite{Voge} and cross section \cite{Beac}. We take from
\cite{Prop} the reactor power and locations. The visible energy
window is chosen to be $E\in [1.22, 7.22]$ MeV (divided in 12
bins) as in \cite{Gouv,GoPe}, with energy resolution
$\sigma(E)/\sqrt{E}=5\%$ ($E$ in MeV) \cite{Shir}. We assume a
``KamLAND year'' corresponding to 550 events/yr for no oscillation
\cite{Shir}, and consider two representative exposure periods of
0.5 and 3 years, the shortest period being a tentative estimate of
the detector live-time that might be used for the first official
release of the KamLAND data.

Concerning the uncertainties, in addition to the statistical
fluctuations in each bin, we include a $\sim4\%$ overall
normalization error \cite{Shir} (due to uncertainties in fiducial
volume, $\nu$ flux, and cross section), and a 2\% energy scale
uncertainty \cite{Shir}. The normalization and energy scale
systematics (fully correlated in each bin) mainly worsen the
reconstruction accuracy of $\sin^2\theta_{12}$ and $\delta m^2$,
respectively. Background reduction and subtraction are still in
progress at KamLAND \cite{Shir}, and we do not include the
(unknown) associated errors in the analysis. More realistic
analyses, including backgrounds and other systematics (e.g., fuel
composition uncertainties \cite{Mura}), will be possible after
release of the real KamLAND data and of related information.

Given the above input, for fixed {\em true\/} values of the
parameters $(\delta m^2,\sin^2\theta_{12})$  we first determine
the $\Delta \chi^2\leq 4$ region including the {\em
reconstructed\/} oscillation parameters, and then project this
region either on $\delta m^2$ or on $\sin^2\theta_{12}$, in order
to obtain the $\pm 2\sigma$ error bands on each of the two
oscillation parameters separately. For simplicity, this procedure
is done for fixed $\theta_{13}=0$. When real KamLAND data will be
available, a more complete analysis procedure (not simulated here)
should include a variable $\theta_{13}$ parameter, to be
constrained by world neutrino data (as previously done for the
currently available data in Figs~3 and 4). Small variations of
$\theta_{13}$ would then basically act as an additional KamLAND
normalization uncertainty \cite{GoPe}.

Figure~6 shows the $\pm2\sigma$ range of the reconstructed $\delta
m^2$, as a function of the true value of $\delta m^2$. The true
value of $\sin^2\theta_{12}$ is fixed at 0.3 (LMA best-fit), while
the reconstructed $\sin^2\theta_{12}$ is projected away. It can be
seen that, in the range most relevant for the LMA solution
($\delta m^2\sim \mathrm{few}\times 10^{-5}$ eV$^2$) the
reconstruction accuracy depends only weakly on $\delta m^2$. It
become rapidly worse, however, outside this range. E.g., for low
values of $\delta m^2$ (approaching the no-oscillation limit in
KamLAND), the reconstructed value of $\delta m^2$ can be
considerably higher than the true one, the variation being
compensated by a lower value of the reconstructed
$\sin^2\theta_{12}$. For high values of $\delta m^2$ (say,
$\gtrsim 10^{-4}$ eV$^2$), degenerate solutions begin to appear,
eventually leading to a tower of (merging) multiple ranges for the
reconstructed $\delta m^2$. For a half-year exposure, an isolated,
low-$\delta m^2$ reconstructed range also appears, since the
spectrum suppression due to true values $\delta m^2\gtrsim
10^{-4}$ eV$^2$ can be confused with a similar suppression induced
by lower values of both $\delta m^2$ {\em and\/}
$\sin^2\theta_{12}$. Such ambiguities can be cleared up, e.g.,
after 3 years of data taking, provided that $\delta m^2\lesssim
2\times 10^{-4}$ eV$^2$. Above this range, KamLAND cannot resolve
the pattern of fast oscillations in the $\overline\nu_e$ energy
spectrum, which could be disentangled only by reducing the typical
baseline \cite{StVi,Piai,HLMA}.

Figure~7 zooms in on true $\delta m^2$ values below $2\times
10^{-4}$ eV$^2$, for three representative (true) values of
$\sin^2\theta_{12}$. For oscillation parameters close to the LMA
ones, the accuracy in the $\delta m^2$ reconstruction can already
be as good as $\sim \pm 10\%$ at $2\sigma$ after half-year
exposure, and can improve by a factor of $\sim 2$ after three
years. The $\delta m^2$ reconstruction accuracy improves (worsens)
slightly for increasing (decreasing) values of
$\sin^2\theta_{12}$, since this parameter governs the amplitude of
the oscillation pattern, from which $\delta m^2$ is inferred.

Finally, Fig.~8 shows the reconstructed range of
$\sin^2\theta_{12}$ (first octant only), for the same choice of
true oscillation parameters as in Fig.~7. From this figure it
appears that the most accurate reconstruction of
$\sin^2\theta_{12}$ occurs for $\delta m^2$ values slightly below
the current LMA best-fit point. The reason is that, for such
relatively low values, the overall rate suppression in KamLAND is
maximized, making it easier to put bounds on the oscillation
amplitude and thus on $\sin^2\theta_{12}$. The $\sin^2\theta_{12}$
reconstruction accuracy tends to become constant for high values
of $\delta m^2$ (regime of fast oscillations). In the lower panel,
notice the appearance of a degenerate reconstructed range for the
case of 0.5 year exposure, which has the same origin as the
corresponding one discussed in Fig.~6.

In conclusion, KamLAND is in an extremely favorable situation to
fix the most likely values of $\delta m^2$ in the LMA region
within a few~\% error. The achievable accuracy in the
$\sin^2\theta_{12}$ reconstruction is instead both lower and more
uncertain, and will depend very much on how the real KamLAND data
will look like.

\section{Conclusions}
We have studied the $3\nu$ perturbations to the usual $2\nu$
solutions to the solar neutrino problem, including---besides the
full set of current solar data---the terrestrial neutrino
constraints coming from the CHOOZ (reactor), SK (atmospheric) and
K2K (accelerator) experiments. The global best fit is reached in
the so-called LMA solution, for the subcase of $2\nu$ mixing
($\theta_{13}=0$). The likelihood of the (less favored) LOW and
QVO solutions does not improve significantly for nonzero
$\theta_{13}$, within the  severe constraints placed by
terrestrial neutrino experiments on such mixing angle. The
anticorrelation between the upper bounds on $\delta m^2$ and
$\theta_{13}$ has been discussed, together with its implications
for the maximum allowed amplitude of possible CP violation effects
(expressed in terms of a representative vacuum-matter invariant).
We have also discussed the expected accuracy of the mass-mixing
parameter reconstruction in the (currently running) KamLAND
experiments, described through a continuous scan of the $\delta
m^2$ parameter within the LMA solution.

\acknowledgments

E.L.\ thanks the organizers of the Summer Institute ``New
Dimensions in Astroparticle Physics'' (Gran Sasso National
Laboratory, Assergi, Italy, 2002), where preliminary results of
this work were presented, for kind hospitality. This work was
supported in part by INFN and in part by the Italian {\em
Ministero dell'Istruzione, Universit\`a e Ricerca\/} through the
``Astroparticle Physics'' research project.



\begin{figure}
\vspace*{1cm}
\includegraphics[scale=0.9, bb= 100 100 500 720]{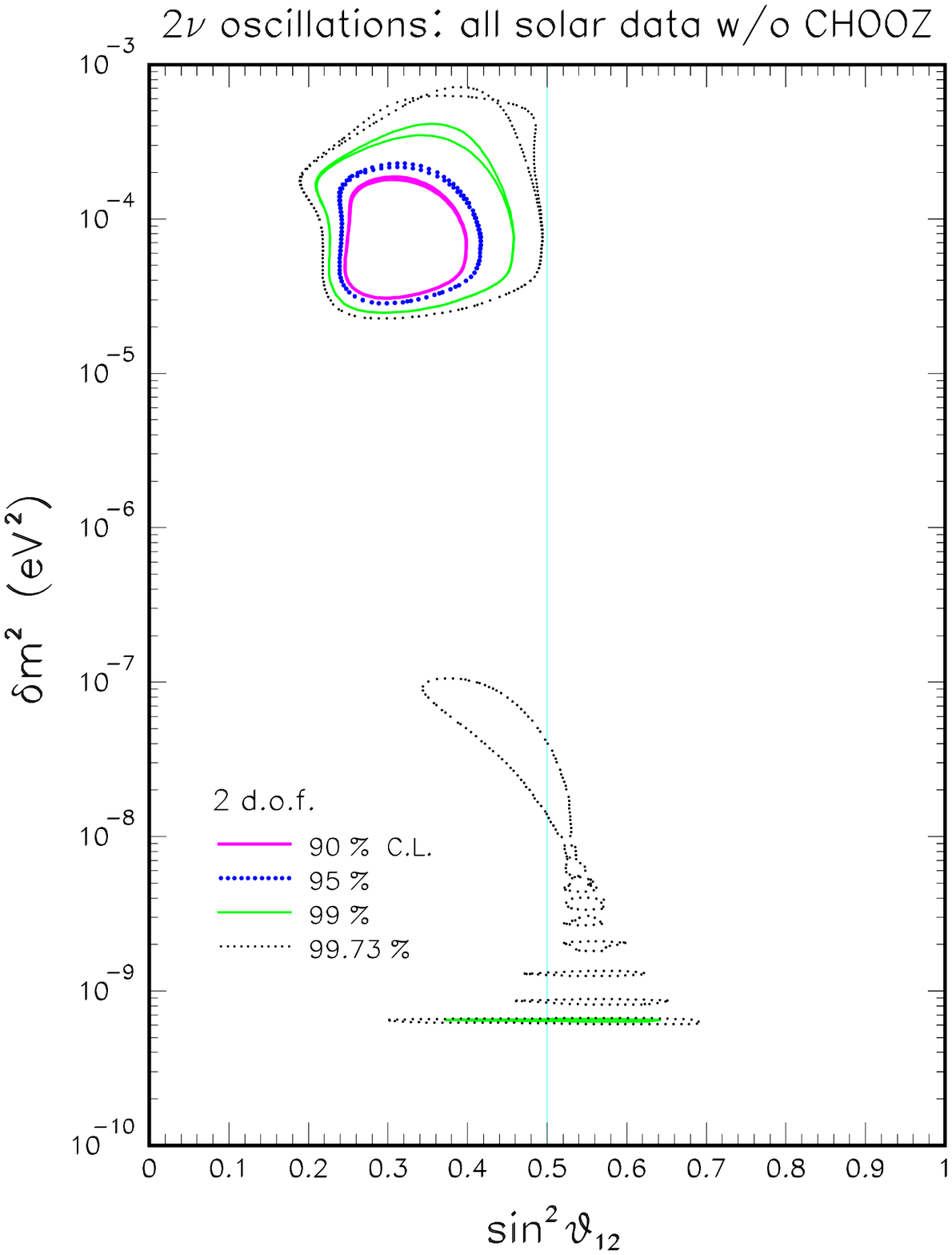}
\vspace*{+0cm} \caption{\label{fig1} Two-flavor global analysis of
solar neutrino oscillations in the $(\delta
m^2,\sin^2\theta_{12})$ parameter space, with and without the
additional constraints placed by the CHOOZ reactor experiment. The
inclusion of CHOOZ leads to slightly more restrictive upper bounds
on $\delta m^2$.}
\end{figure}

\begin{figure}
\vspace*{1cm}
\includegraphics[scale=0.9, bb= 100 100 500 720]{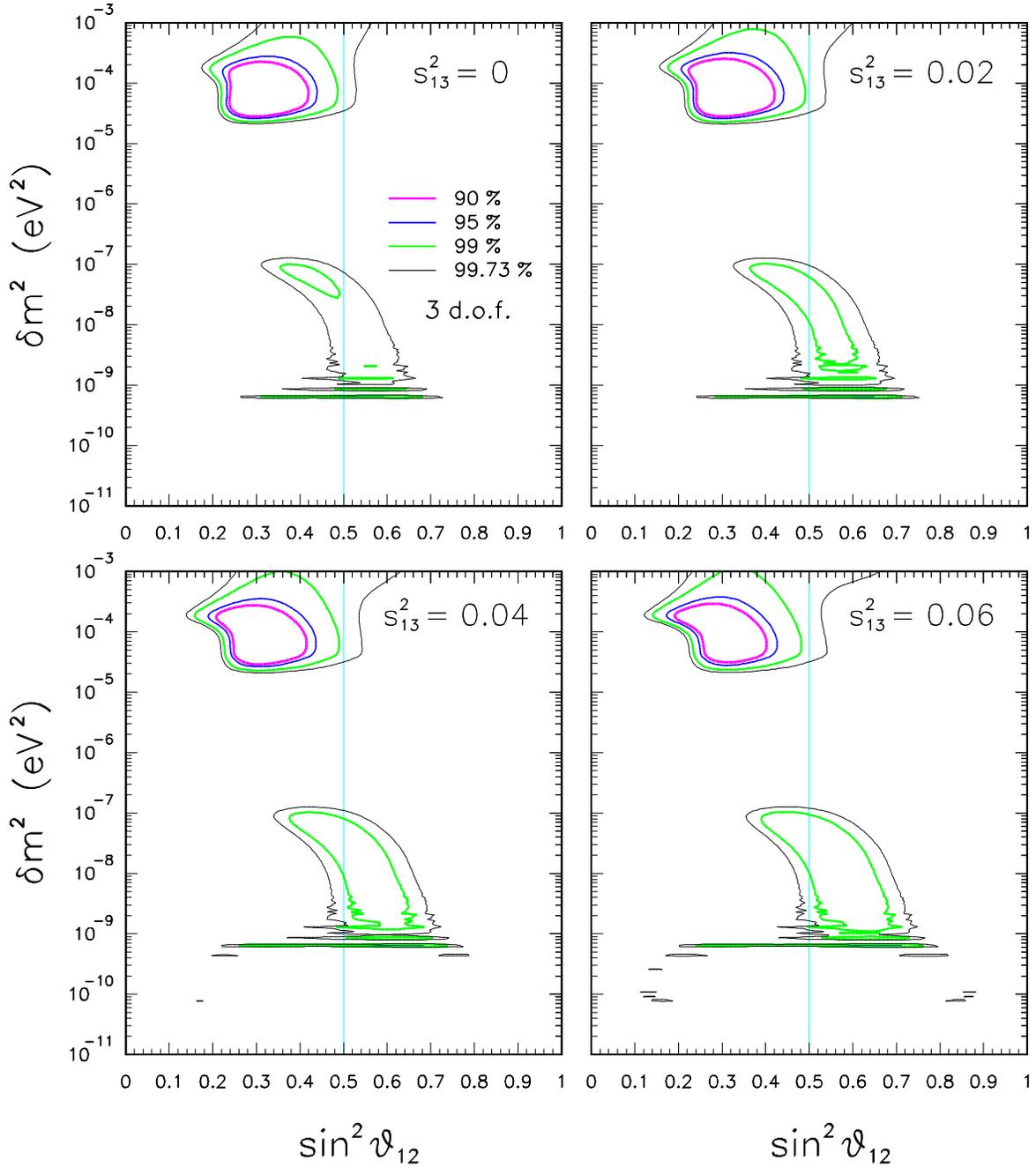}
\vspace*{+0cm} \caption{\label{fig2} Three-flavor global analysis
of solar neutrino oscillations in the $(\delta
m^2,\sin^2\theta_{12},\sin^2\theta_{13})$ parameter space, shown
through four sections at fixed values of $s^2_{13}\equiv
\sin^2\theta_{13}$.}
\end{figure}

\begin{figure}
\vspace*{1cm}
\includegraphics[scale=0.9, bb= 100 100 500 720]{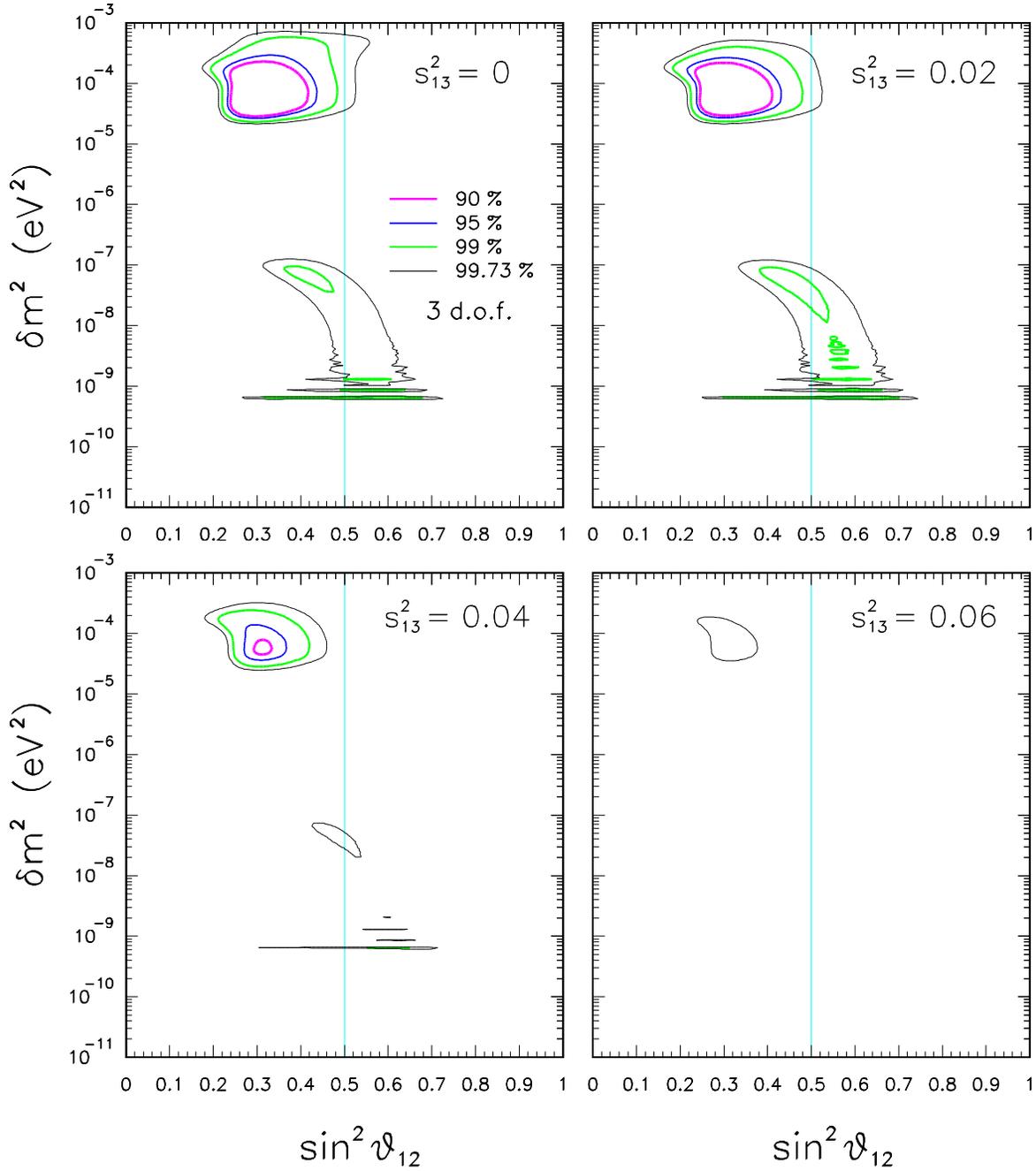}
\vspace*{+0cm} \caption{\label{fig3} As in Fig.~\ref{fig2}, but
including the constraints from terrestrial neutrino oscillation
searches at CHOOZ, SK, and K2K.}
\end{figure}

\begin{figure}
\includegraphics[scale=0.9, bb= 100 100 500 720]{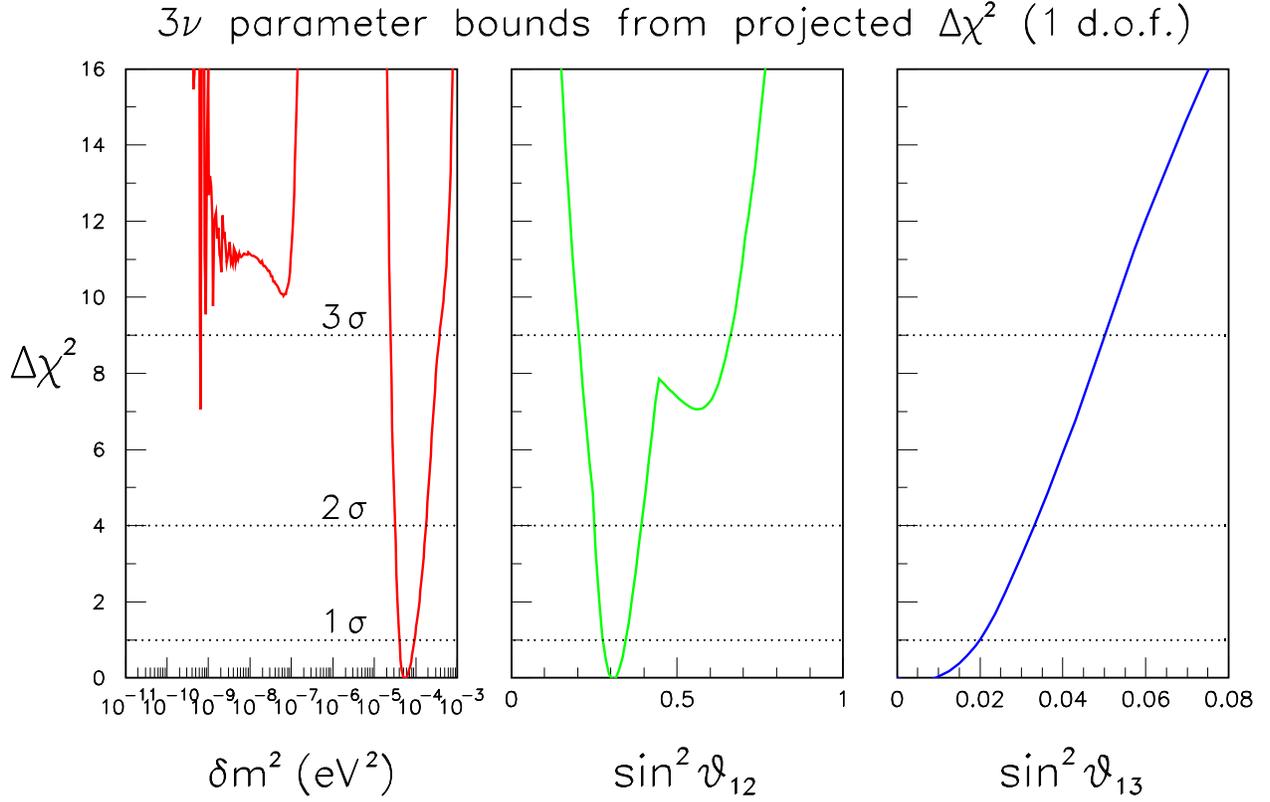}
\vspace*{-2cm} \caption{\label{fig4} Projections of the global
(solar+terrestrial) $\Delta\chi^2$ function onto each of the
$(\delta m^2,\sin^2\theta_{12},\sin^2\theta_{13})$ parameters. The
$n$-sigma bounds on each parameter (the others being
unconstrained) correspond to $\Delta\chi^2=n^2$.}
\end{figure}

\begin{figure}
\vspace*{1cm}
\includegraphics[scale=0.9, bb= 100 100 500 720]{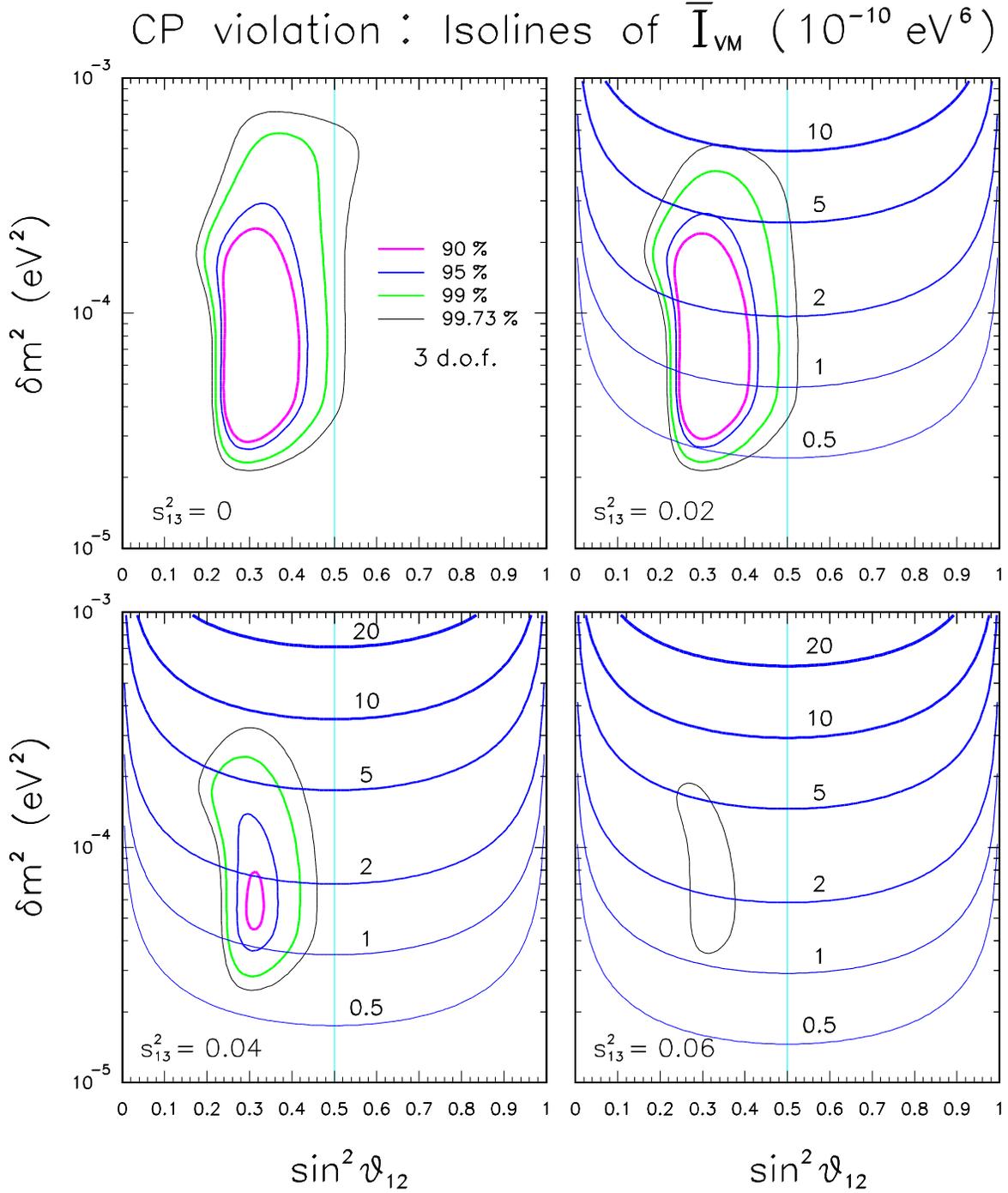}
\vspace*{+0cm} \caption{\label{fig5} Isolines of the vacuum-matter
invariant ${\overline I}_\mathrm{VM}$, in units of $10^{-10}$
eV$^6$. Superposed are the contours of the large mixing angle
solution. The invariant ${\overline I}_\mathrm{VM}$ appears as a
typical prefactor in  CP violating amplitudes expected at future
accelerator searches; see the text for details. }
\end{figure}

\begin{figure}
\vspace*{-5cm}
\includegraphics[scale=1.0, bb= 30 100 500 700]{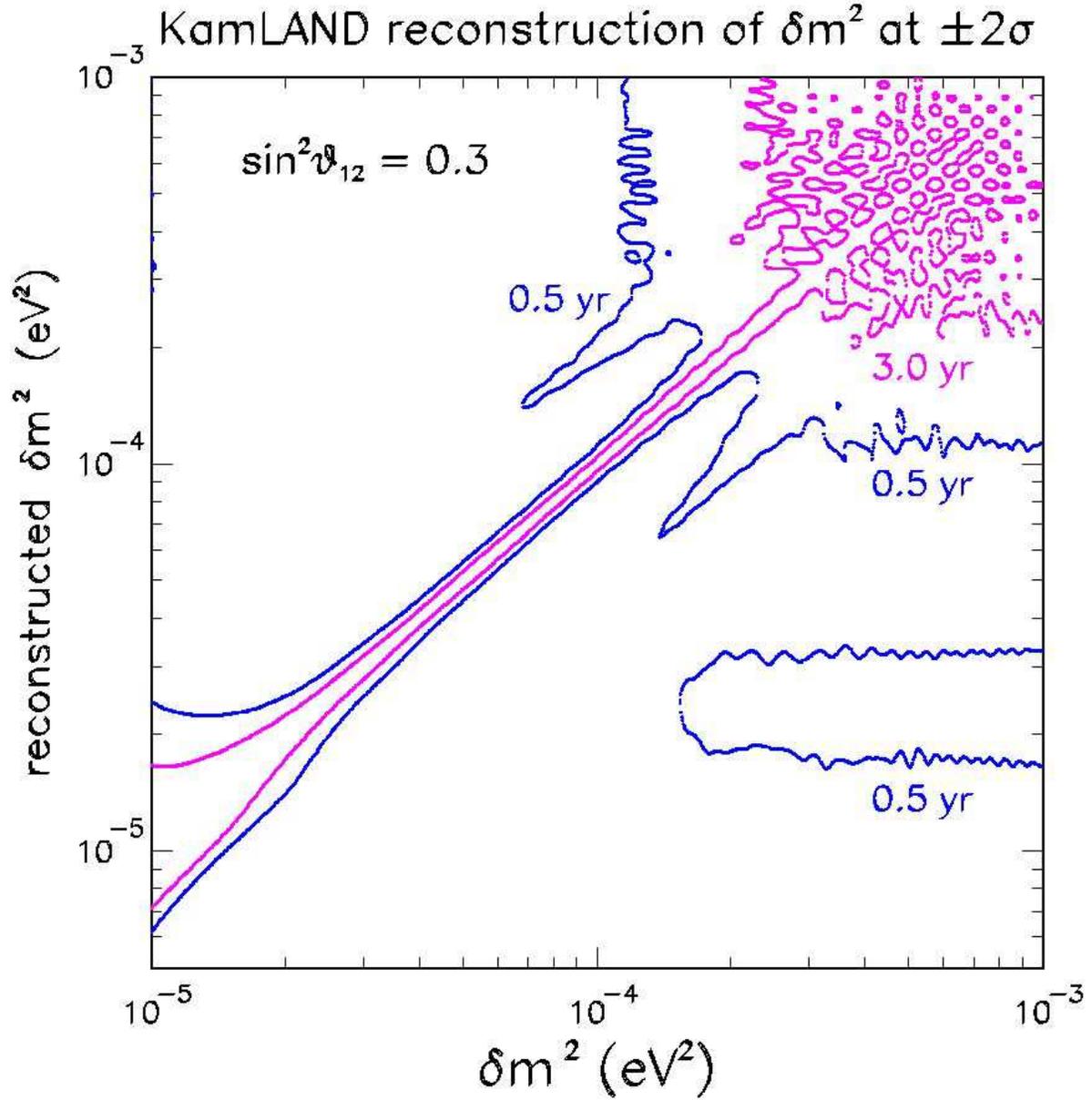}
\vspace*{+3cm} \caption{\label{fig6} Simulation of KamLAND
results. Reconstructed range of $\delta m^2$ at $\pm 2\sigma$, as
a function of the true value of $\delta m^2$ in abscissa. The true
value of $\sin^2\theta_{12}$ is fixed at 0.3. The curves refer to
detector exposures of 0.5 and 3 years.}
\end{figure}

\begin{figure}
\vspace*{-5cm}
\includegraphics[scale=1.0, bb= 30 100 500 700]{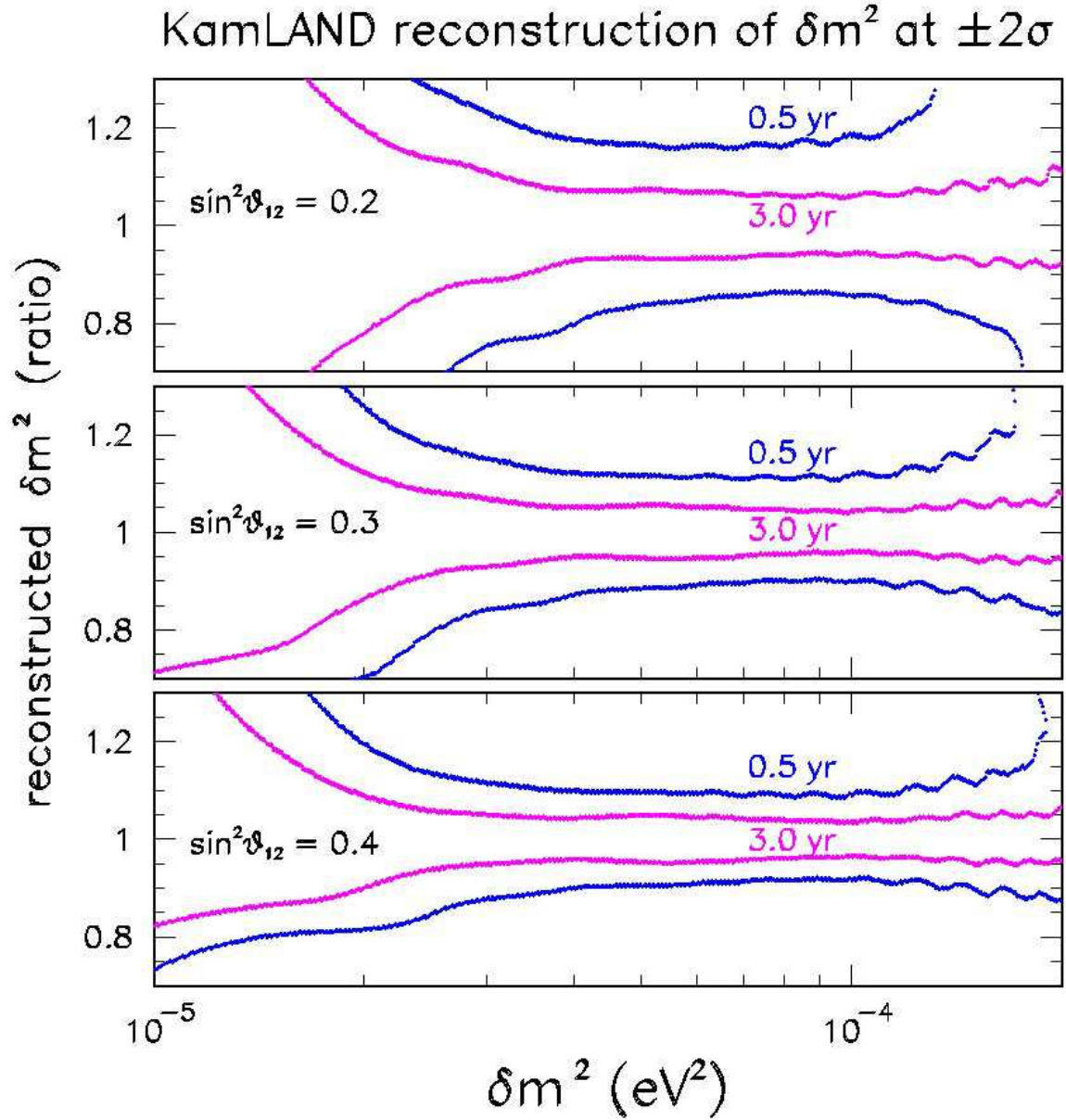}
\vspace*{+3cm} \caption{\label{fig7} Simulation of KamLAND
results. Reconstructed range of $\delta m^2$ at $\pm 2\sigma$
(normalized to the true value of $\delta m^2$ in abscissa), for
three representative values of $\sin^2\theta_{12}$, and for
exposures of 0.5 and 3 years. }
\end{figure}

\begin{figure}
\vspace*{-5cm}
\includegraphics[scale=1.0, bb= 30 100 500 700]{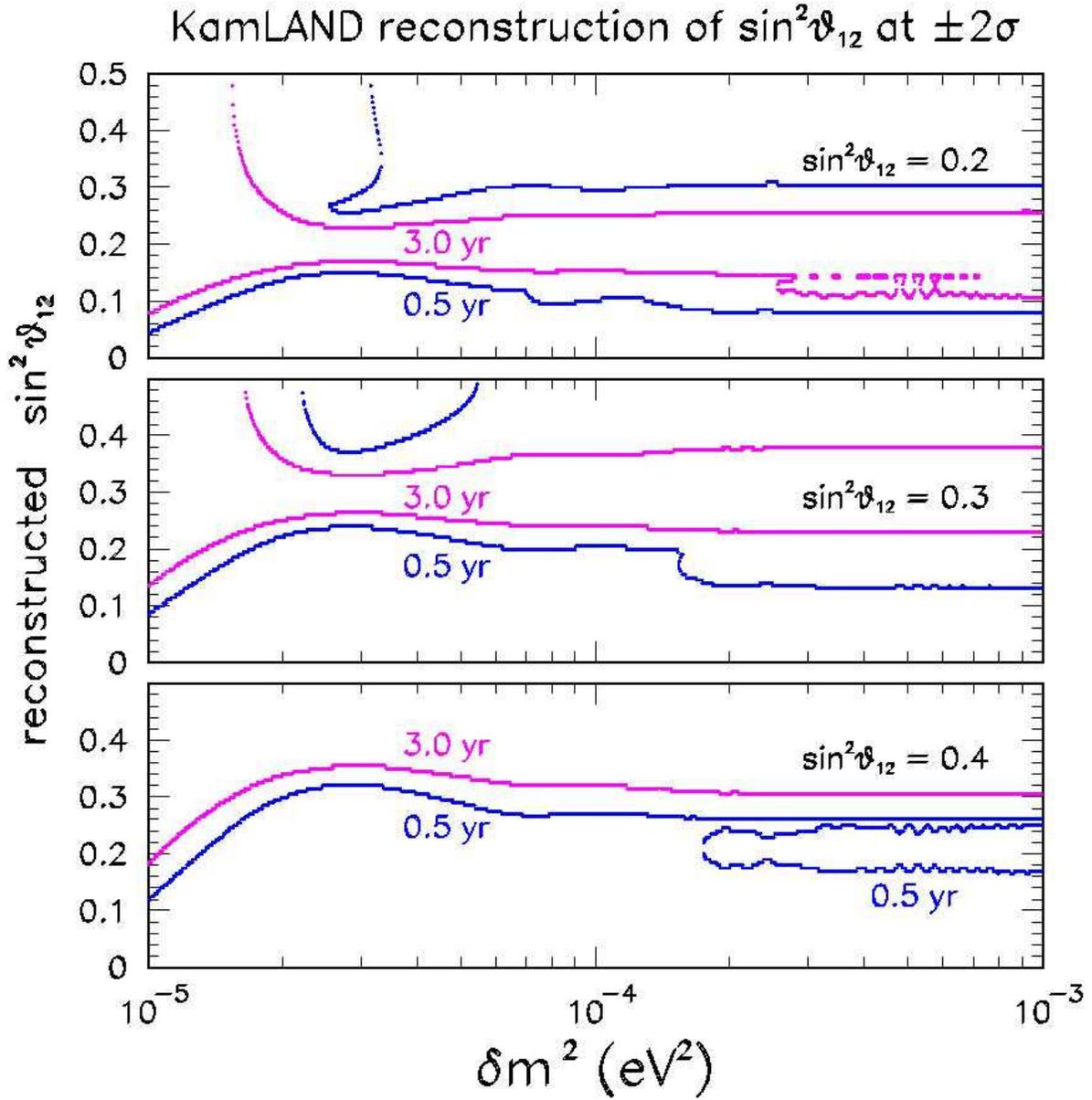}
\vspace*{+3cm} \caption{\label{fig8} Simulation of KamLAND
results. Reconstructed range of $\sin^2\theta_{12}$ at $\pm
2\sigma$ as a function of $\delta m^2$, for three representative
(true) values of $\sin^2\theta_{12}$, and for exposures of 0.5 and
3 years. }
\end{figure}

\end{document}